\documentclass{ws-p8-50x6-00}
\newcommand{\beq}{\begin{equation}}
\newcommand{\eeq}{\end{equation}}
\newcommand{\beqa}{\begin{eqnarray}}
\newcommand{\eeqa}{\end{eqnarray}}
\newcommand{\ba}{\begin{array}}
\newcommand{\ea}{\end{array}}

\begin{document}

\title{Anomalous Spectral Statistics 
of the Asymmetric Rotor Model} 

\author{V.R. Manfredi}

\address{Dipartimento di Fisica "G. Galilei", Universit\`a di Padova, \\
Istituto Nazionale di Fisica Nucleare, Sezione di Padova, \\
Via Marzolo 8, 35131 Padova, Italy, \\ 
E-mail: manfredi@pd.infn.it}

\author{L. Salasnich}

\address{Istituto Nazionale per la Fisica della Materia, 
Unit\`a di Milano Universit\`a,\\ 
Dipartimento di Fisica, Universit\`a di Milano, \\ 
Via Celoria 16, 20133 Milano, Italy
\\
E-mail: salasnich@mi.infm.it}  

\maketitle

\abstracts{
We investigate the spectral statistics of the asymmetric rotor model 
(triaxial rigid rotator). 
The asymmetric top is classically integrable and, according 
to the Berry-Tabor theory, its spectral statistics should be 
Poissonian. Surprisingly, our numerical 
results show that the nearest neighbor spacing distribution 
$P(s)$ and the spectral rigidity $\Delta_3(L)$ 
do not follow Poisson statistics.}

\section{Introduction}

As is well known, in the semiclassical limit\cite{r1,r2} there is 
a clear connection between the behavior of classical systems 
(regular or chaotic) and the corresponding quantal ones. 
In particular for quantal systems, corresponding to classical 
regular systems, the spectral statistics ($P(s)$ and $\Delta_3(L)$) 
follow the Poisson ensemble, while for systems corresponding 
to chaotic ones the Wigner ensemble is followed 
(see, for example, Ref. 3 and references therein). 
\par
Nevertheless, some exceptions are known. The most famous case is 
perhaps the harmonic oscillator one, discussed in great detail 
in Ref. 4 and 5. 
\par
The aim of this paper is to discuss another pathological case: 
the classically integrable triaxial rotator model (see, for 
instance, Ref. 6). Incidentally, this model has been used 
very often in the description of the low-lying states of the 
even-even atomic nuclei.\cite{r7} 
\par 
The asymmetric top described by the rotor model 
is a classically integrable system, but 
an analytical formula for its quantum energy spectrum it not known. 
Nevertheless, numerical results can be obtained. 
By following the Landau approach, the Hamiltonian operator 
is split into 4 submatrices, corresponding to different symmetry classes. 
Each truncated submatrix is numerically diagonalized. 
Finally, the nearest-neighbor spacing distribution $P(s)$ 
and the spectral rigidity $\Delta_3(L)$ are calculated. 
Surprisingly, the spectral statistics of energy levels do not 
follow the predictions of Poisson statistics. 
This result may suggest the presence of a hidden symmetry 
in the system but, to our knowledge, such a symmetry 
has not yet been identified. 

\section{The Asymmetric Rotor Model}

Let us consider a system of coordinates with axes along 
the three principal axes of intertia of the top, and rotating 
with it. The classical Hamiltonian $H$ of the top is given by 
\beq 
{\hat H} = {1 \over 2} 
\left( a J_1^2 + b J_2^2  + c J_3^2 \right) \; , 
\eeq 
where ${\vec J}=(J_1,J_2,J_3)$ is the angular momentum of the 
rotation and $a=1/I_1$, $b=1/I_2$, $c=1/I_3$ are three 
parameters such that $I_1$, $I_2$ and $I_3$ are 
the principal momenta of intertia of the top. 
The Hamiltonian is classically integrable 
and its action variables are precisely 
the three components $J_s$, $s=1,2,3$, of the angular momentum. 
\par 
The quantum Hamiltonian ${\hat H}$ is 
obtained by replacing the components of the 
angular momentum, in the classical expression 
of the energy, by the corresponding quantum operators 
${\hat J}_1$, ${\hat J}_2$ and ${\hat J}_3$.  
The commutation rules for the operators of the angular momentum 
components in the rotating system of coordinates are given by 
\beq 
{\hat J}_r {\hat J}_s - {\hat J}_s {\hat J}_r 
= - i\hbar \; \epsilon_{rst} \; {\hat J}_t \; , 
\eeq 
where $\epsilon_{rst}$ is the Ricci tensor and $r,s,t=1,2,3$. 
Note that these commutation rules  
differ from those in the fixed system in the sign on the 
right-hand side.\cite{r8} 
\par 
As usual, the two operators 
${\hat J}^2={\hat J}_1^2+{\hat J}_2^2+{\hat J}_3^2$ 
and ${\hat J}_3$ are simultaneously diagonalized on the basis of 
eigenstates $|J,k\rangle$ with integer eigenvalues $J$ and $k$ 
($k=-J,-J+1,...,J-1,J$), respectively. 
The non-zero matrix elements of the quantum Hamiltonian ${\hat H}$ 
in the basis $|J,k\rangle$ are given by 
\beq 
\langle J,k|{\hat H}|J,k \rangle = {\hbar^2 \over 4} (a+b) 
(J(J+1)-k^2) + {\hbar^2 \over 2} c k^2 \; , 
\eeq
$$
\langle J,k|{\hat H}|J,k+2 \rangle = \langle J,k+2|H|J,k \rangle = 
$$ 
\beq
= {\hbar^2 \over 8} (a-b)\sqrt{ (J-k)(J-k-1)(J+k+1)(J+k+2) } \; . 
\eeq 
The quantum Hamiltonian ${\hat H}$ has matrix elements only 
for transitions with $k\to k$ or $k\pm2$. The absence of matrix 
elements for transitions between states with even and odd $k$ 
has the result that the matrix of degree $2J+1$ is the 
direct product of two matrices of degrees 
$J$ and $J+1$. One of these contains matrix elements for transitions 
between states with even $k$, and the other contains those 
for transitions between states with odd $k$.\cite{r8} 
\par 
It is useful to introduce a new basis, given by 
\par 
$$ 
|J,k;S\rangle 
= {1\over \sqrt{2}} \left( |J,k\rangle + |J,-k\rangle \right) \; , 
\;\;\;\;\; |J,0,S\rangle =  |J,0\rangle \; , 
$$ 
\beq 
|J,k;A\rangle 
= {1\over \sqrt{2}} \left( |J,k\rangle - |J,-k\rangle \right) 
\; , \;\;\;\;\; k\neq 0 \; . 
\eeq 
By using this new basis, the total Hamiltonian matrix 
is decomposed in the direct product of $4$ submatrices 
by considering the parity of the quantum 
number $k$: even (E) or odd (O), and the symmetry of the state: 
symmetric (S) or anti-symmetric (A). 
So the submatrices are labelled as follow: (E,S), (E,A), (O,S), (O,A). 
These are the classes of symmetry of the system. 
In Table 1 we show the dimension of each submatrix for a fixed $J$. 

\begin{table}[t]
\caption{ Number of states in each submatrix of the 
asymmetrical top Hamiltonian for a fixed $J$.} 
\begin{center}
\begin{tabular}{|ccccc|} \hline 
$$ & $(E,S)$ & $(E,A)$ & $(O,S)$ & $(O,A)$ \\ 
\hline 
$J$ even & ${J\over 2}+1$ & ${J\over 2}$   & ${J\over 2}$   
& ${J\over 2}$ \\ 
$J$ odd  & ${J-1\over 2}$ & ${J+1\over 2}$ & ${J+1\over 2}$ 
& ${J+1\over 2}$ \\ 
\hline 
\end{tabular} 
\end{center} 
\end{table}

\par
The matrix elements of the Hamiltonian ${\hat H}$ in the 
new basis, with respect to the old basis, are given by 
\beq
\langle J,k,S |{\hat H}|J,k,S \rangle = 
\langle J,k,A |{\hat H}|J,k,A \rangle 
= \langle J,k|{\hat H}|J,k \rangle 
\; , \;\;\;\;\; k\neq 1 
\eeq
\beq 
\langle J,1,S |{\hat H}|J,1,S \rangle = 
\langle J,1|{\hat H}|J,1 \rangle \; + \; 
\langle J,1|{\hat H}|J,-1\rangle 
\eeq
\beq 
\langle J,1,A |{\hat H}|J,1,A \rangle = 
\langle J,1|{\hat H}|J,1 \rangle \; - \; 
\langle J,1|{\hat H}|J,-1\rangle 
\eeq
\beq
\langle J,k,S |{\hat H}|J,k+2,S \rangle = 
\langle J,k,A |{\hat H}|J,k+2,A \rangle 
= \langle J,k|{\hat H}|J,k+2 \rangle 
\; , \;\;\;\;\; k\neq 0 
\eeq
\beq
\langle J,0,S |{\hat H}|J,2,S \rangle = 
\sqrt{2} \langle J,0|{\hat H}|J,2 \rangle 
\; , \;\;\;\;\; k\neq 0 
\eeq
We calculate the eigenvalues of each submatrix 
for different values of $J$ using a fast implementation, 
in double precision, 
of the Lanczos algorithm with a LAPAC code.\cite{r9} 

\begin{figure}[t]
\epsfxsize=26pc 
\epsfbox{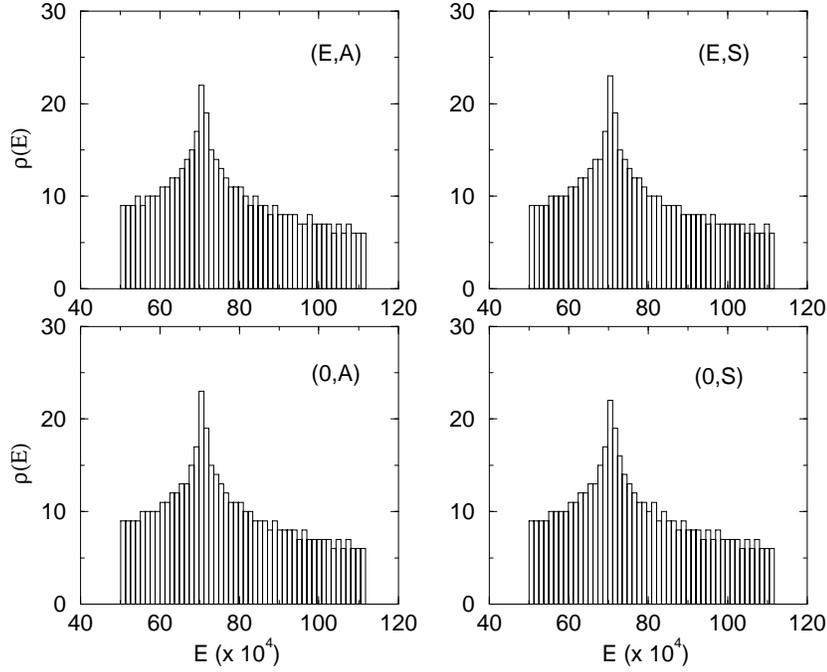} 
\caption{Density of levels $\rho(E)$ of the 
four classes of symmetry with $J=1000$. 
Parameters: $a=1$, $b=\sqrt{2}$, 
$c=\sqrt{5}$ and $\hbar=1$.}
\end{figure} 

In Figure 1 we plot the density of levels $\rho(E)$ 
of each submatrix of ${\hat H}$ and $J=1000$. 
The results show that the density of levels is practically 
the same for the four classes. $\rho(E)$ displays a high peak 
at the left-center of the energy interval and a long tail 
for large energy values. 

\section{Spectral Statistics} 

As previously discussed, according to the 
Berry-Tabor theory\cite{r10,r11}, 
given a classical integrable Hamiltonian that, 
written in action variables $J_r$, 
satisfies the condition 
$\left| {\partial^2 H\over \partial J_r \partial J_s} 
\right| \neq 0$, then, in the semiclassical limit, 
its spectral statistics should follow the Poisson statistics. 
Note that a system of linear harmonic oscillators, whose 
Hamiltonian is given by $H={\vec \omega}\cdot {\vec I}$, 
does not satisfy the previous condition. 
In fact, a system of linear harmonic oscillators is integrable 
but it does not follow Poissonian statistics.\cite{r4,r5} 
\par 
The triaxial rigid rotator is integrable and satisfies 
the previous Berry-Tabor condition. Thus, one expects that 
the spectral statistics of the quantized rigid rotator 
should be Poissonian. We shall show that is not the case. 

\begin{figure}[t]
\epsfxsize=26pc 
\epsfbox{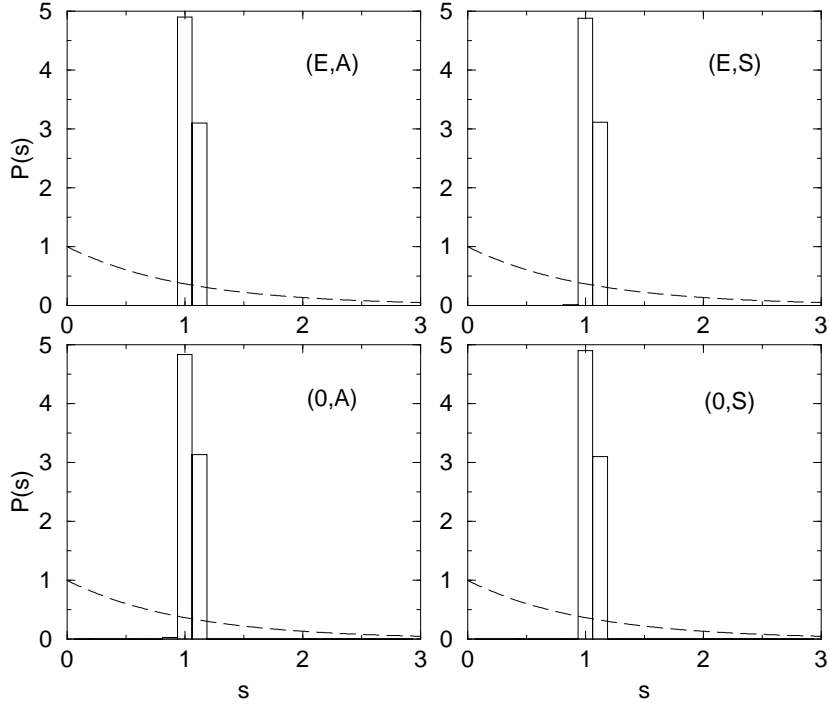} 
\caption{Nearest neighbor spacing distribution 
$P(s)$ of the four classes of symmetry with $J=1000$. 
The dashed line is the Poisson prediction $P(s)=\exp{(-s)}$. 
Parameters: $a=1$, $b=\sqrt{2}$, 
$c=\sqrt{5}$ and $\hbar=1$.}
\end{figure}

\par 
In general, various statistics may be used to show the local correlations 
of the energy levels but the most used spectral statistics 
are $P(s)$ and $\Delta_3(L)$. $P(s)$ is 
the distribution of nearest-neighbor spacings 
$s_i=({\tilde E}_{i+1}-{\tilde E}_i)$ 
of the unfolded levels ${\tilde E}_i$. 
It is obtained by accumulating the number of spacings that lie within 
the bin $(s,s+\Delta s)$ and then normalizing $P(s)$ to unit. 
As shown by Berry and Tabor\cite{r10,r11}, 
for quantum systems whose classical analogs are integrable, 
$P(s)$ is expected to follow the Poisson distribution 
\beq 
P(s)=\exp{(-s)} \; . 
\eeq 

\begin{figure}[t]
\epsfxsize=26pc 
\epsfbox{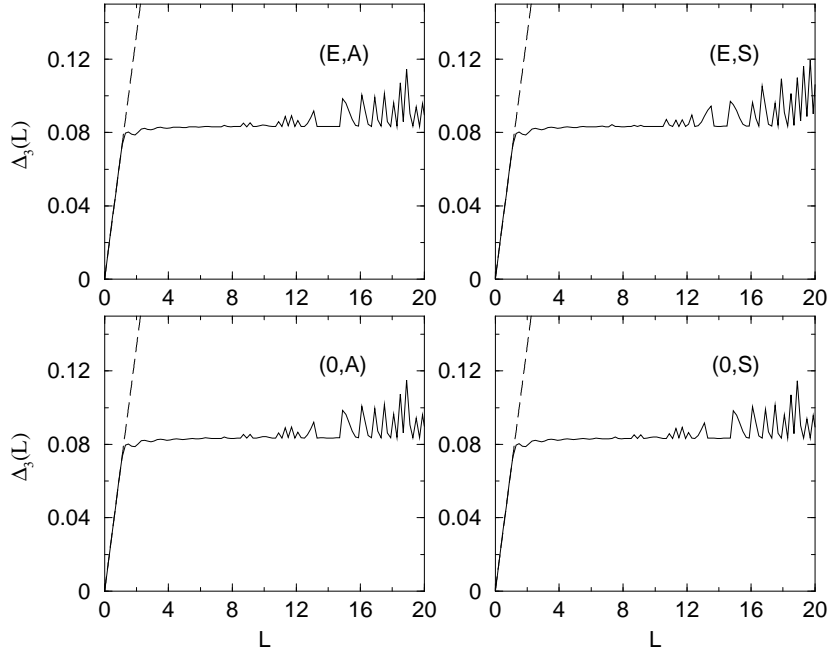} 
\caption{Spectral rigidity $\Delta_3(L)$ of the 
four classes of symmetry with $J=1000$. 
The dashed line is the Poisson prediction $\Delta_3(L)=L/15$. 
Parameters: $a=1$, $b=\sqrt{2}$, $c=\sqrt{5}$ and $\hbar=1$.}
\end{figure}

The statistic $\Delta_3(L)$ is defined, for a fixed interval 
$(-L/2,L/2)$, as the least-square deviation of the staircase 
function $N(E)$ from the best straight line fitting it: 
$$
\Delta_{3}(L)={1\over L}\min_{A,B}\int_{-L/2}^{L/2}[N(E)-AE-B]^2 dE \; , 
$$
where $N(E)$ is the number of levels between E and zero for positive 
energy, between $-E$ and zero for negative energy. The $\Delta_{3}(L)$ 
statistic provides a measure of the degree of rigidity of the 
spectrum: for a given interval L, the smaller $\Delta_{3}(L)$ is, 
the stronger is the rigidity, signifying the long-range 
correlations between levels. For this statistic 
the Poissonian prediction is 
\beq 
\Delta_3(L)= {L\over 15} \; . 
\eeq 
It is useful to remember that Berry, on the basis 
of the Gutwiller semiclassical formula for the density of states, 
has shown that $\Delta_3(L)$ deviates from the universal 
Poissonian predictions for large $L$: 
$\Delta_3(L)$ should saturate to an asymptotic value 
performing damped oscillations.\cite{r12} 
\par 
In Figure 2 the spectral statistic 
$P(s)$ is plotted for the four 
submatrices of ${\hat H}$ and $J=1000$.  
Note that the level spectrum is mapped into unfolded levels 
with quasi-uniform level density 
by using a standard procedure described in Ref. 13. 
As expected from the previous analysis of density of levels, 
$P(s)$ is practically the same for the four classes of symmetry. 
Moreover, $P(s)$ has a pathological behavior: a peak near 
$s=1$ and nothing elsewhere. 
\par
Compared to $P(s)$, the spectral rigidity $\Delta_3(L)$ 
is less pathological. As shown in Figure 3, $\Delta_3(L)$ 
follows quite well the Poisson prediction $\Delta_3(L)=L/15$ 
for small $L$ but for larger values of $L$ 
it gets a constant mean value with 
fluctuations around this mean value. These fluctuations 
becomes very large by increasing $L$, in contrast 
with the Berry prediction\cite{r12}. 
Note that we have repeated the caculations 
with other values of $a$, $b$ and $c$ but the results 
do not change. 

\begin{figure}[t]
\epsfxsize=26pc 
\epsfbox{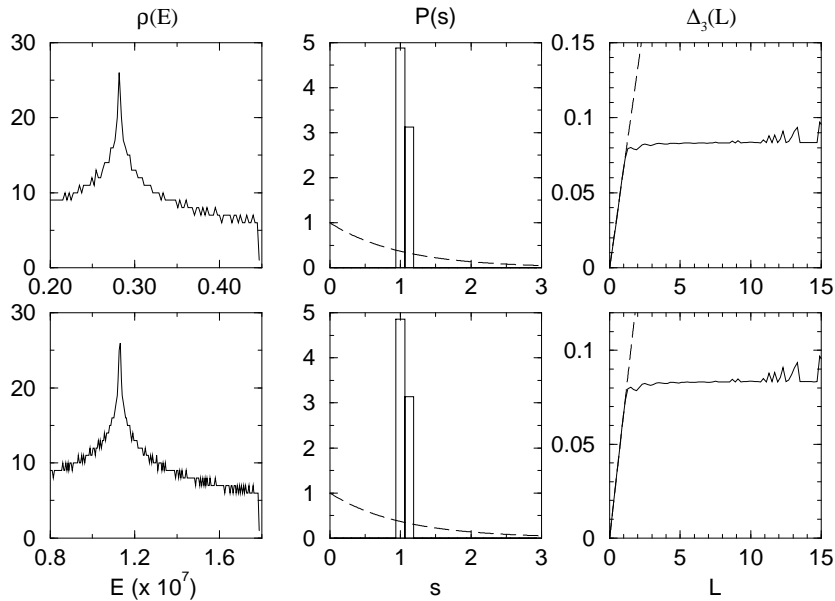} 
\caption{Density of levels $\rho(E)$, 
Nearest neighbor spacing distribution 
$P(s)$ and spectral rigidity $\Delta_3(L)$ 
of the $(E,S)$ class of symmetry, with 
$J=2000$ (top) and $J=4000$ (bottom). 
Dashed lines are Poisson predictions. 
Parameters: $a=1$, $b=\sqrt{2}$, 
$c=\sqrt{5}$ and $\hbar=1$.}
\end{figure}

\par 
The behavior of the density of levels $\rho(E)$ and 
of the spectral statistics $P(s)$ and $\Delta_3(L)$ 
does not change by changing the matrix dimension, 
namely the quantum number $J$. In Figure 4 we plot 
the density of levels and the spectral statistics 
for $J=2000$ and $J=4000$. 

\section*{Conclusions} 

The main conclusion of this paper 
is that the asymmetric rotor is, like the harmonic 
oscillator, another pathological case with respect 
to the classical-quantum correspondence 
between integrability and Poisson statistics.  
In our opinion, the pathology of the asymmetric rotor model 
is more interesting because, unlike the harmonic oscillator, 
the asymmetric rotor satisfies the conditions 
of the Berry-Tabor theory. 
The presence of hidden symmetries could explain 
the pathological behavior of spectral statistics 
but such symmetries have not yet been identified. 

\vskip 0.5cm 

V.R.M. is greately indebted to Y. Alhassid, M.V. Berry, S. Graffi 
and V. Zelevinsky, for enlightening discussions. 
L.S. acknowledges M. Robnik for fruitful conversations.


\begin{thebibliography}{99}

\bibitem{r1} V.P. Maslov and M.V. Fedoriuk, {\it Semiclassical Approximation 
in Quantum Mechanics} (Reidel Publishing Company, 1981). 

\bibitem{r2} A.B. Migdal, {\it Qualitative Methods in Quantum Theory}  
(Benjamin, 1997). 

\bibitem{r3} V.R. Manfredi and L. Salasnich, 
Int. J. Mod. Phys. B {\bf 13}, 2343 (1999). 

\bibitem{r4} A. Pandey, O. Bohigas and M.J. Giannoni: J. Phys. A: Math. Gen. 
{\bf 22}, 4083 (1989). 

\bibitem{r5} A. Pandey and R. Ramaswamy: Phys. Rev. A {\bf 43}, 4237 (1991). 

\bibitem{r6} H. Goldstein, {\it Classical Mechanics} 
(Addision Wesley, Reading, 1980). 

\bibitem{r7} J.M. Eisenberg and W. Greiner, {\it Nuclear Models}, vol. 1 
(North Hollnd, Amsterdam, 1975). 

\bibitem{r8} L. Landau and E. Lifshitz, {\it Course in Theoretical Physics}, 
vol. 3: Quantum Mechanics (Pergamon, 1977).  

\bibitem{r9} LAPAC Fortran Library, Linear Algebra Package, NAG Ltd 2001. 

\bibitem{r10} M.V. Berry and M. Tabor, 
Proc. Roy. Soc. Lond. A {\bf 356}, 375 (1977); 
M.V. Berry, Annals of Phys. {\bf 131}, 163 (1981). 

\bibitem{r11} M. Tabor {\it Chaos and Integrability in Nonlinear Dynamics} 
(Wiley, New York, 1989). 

\bibitem{r12} M. Berry, Proc. Roy. Soc. Lond. A {\bf 400}, 229 (1985). 

\bibitem{r13} V.R. Manfredi, Lett. Nuovo Cimento {\bf 40}, 135 (1984). 

\end{thebibliography}
\end{document}